\documentclass[11pt]{article}
\usepackage{amsmath,amssymb,amsthm,amsxtra,overpic,bbm,bm,epsfig,ulem,color,multirow}
\usepackage{braket}

\begin{document}

\begin{center}
{\Large Consequences of the $\mu$-$\tau$ reflection symmetry for leptogenesis
\\ in a seesaw model with diagonal Dirac neutrino mass matrix }
\end{center}

\vspace{0.05cm}

\begin{center}
{\bf Zhen-hua Zhao\footnote{zhaozhenhua@lnnu.edu.cn} , \bf Yan Shao, \bf Hong-Yu Shi } \\
{ $^1$ Department of Physics, Liaoning Normal University, Dalian 116029, China \\
$^2$ Center for Theoretical and Experimental High Energy Physics, \\ Liaoning Normal University, Dalian 116029, China }
\end{center}

\vspace{0.2cm}

\begin{abstract}
In this paper, we have studied the consequences of the $\mu$-$\tau$ reflection symmetry for leptogenesis in the type-I seesaw model with diagonal Dirac neutrino mass matrix. We have first obtained the phenomenologically allowed values of the model parameters, which show that there may exist zero or equal entries in the Majorana mass matrix for the right-handed neutrinos, and then studied their predictions for three right-handed neutrino masses, which show that there may exist two nearly degenerate right-handed neutrinos. Then, we have studied the consequences of the model for leptogenesis. Due to the $\mu$-$\tau$ reflection symmetry, leptogenesis can only work in the two-flavor regime. Furthermore, leptogenesis cannot work for the particular case of $r=1$. Accordingly, for some benchmark values of $r \neq 1$, we have given the constraints of leptogenesis for relevant parameters.
Furthermore, we have investigated the possibilities of leptogenesis being induced by the renormalization group evolution effects for two particular scenarios. For the particular case of $r=1$, the renormalization group evolution effects will break the orthogonality relations among different columns of $M^{}_{\rm D}$ and consequently induce leptogenesis to work. For the low-scale resonant leptogenesis scenario which is realized for nearly degenerate right-handed neutrinos, the renormalization group evolution effects can break the $\mu$-$\tau$ reflection symmetry and consequently induce leptogenesis to work.
\end{abstract}

\newpage

\section{Introduction}

As we know, the phenomenon of neutrino oscillations reveals that neutrinos are massive and different lepton flavors are mixed \cite{xing}. In order to accommodate such a phenomenon beyond the Standard Model (SM), one must extend the SM (where neutrinos are massless) in a proper way. In the literature, one of the most popular and natural ways of generating the nonzero but tiny neutrino masses is the type-I seesaw model which introduces super heavy right-handed neutrinos $N^{}_I$ (for $I=1, 2, 3$) into the SM \cite{seesaw}. Like other fermions, the right-handed neutrinos can constitute the Yukawa coupling operators with the left-handed neutrinos $\nu^{}_\alpha$ (for $\alpha = e, \mu, \tau$) which reside in the lepton doublets $L^{}_\alpha$ and the Higgs doublet $H$: $(Y^{}_{\nu})^{}_{\alpha I} \overline {L^{}_\alpha} H N^{}_I $ with $(Y^{}_{\nu})^{}_{\alpha I}$ being the Yukawa coupling coefficients. These operators will contribute the Dirac neutrino mass terms $(M^{}_{\rm D})^{}_{\alpha I}= (Y^{}_{\nu})^{}_{\alpha I} v$ [here $(M^{}_{\rm D})^{}_{\alpha I}$ is the $\alpha I$ element of the Dirac neutrino mass matrix $M^{}_{\rm D}$] after the neutral component of $H$ acquires a nonzero vacuum expectation value $v = 174$ GeV. What is different is that the right-handed neutrinos themselves can also have the Majorana mass terms $\overline{N^c_I} (M^{}_{\rm R})^{}_{IJ} N^{}_J$ with the superscript $c$ denoting the charge conjugation of relevant fields [here $(M^{}_{\rm R})^{}_{IJ}$ is the $IJ$ element of the right-handed neutrino mass matrix $M^{}_{\rm R}$]. Then, under the seesaw condition $M^{}_{\rm R} \gg M^{}_{\rm D}$, one will obtain an effective Majorana mass matrix as
\begin{eqnarray}
M^{}_{\nu} = - M^{}_{\rm D} M^{-1}_{\rm R} M^{T}_{\rm D} \;,
\label{1}
\end{eqnarray}
for three light neutrinos by integrating the heavy right-handed neutrinos out. Thanks to such a formula, the smallness of neutrino masses can be naturally explained by the heaviness of right-handed neutrinos.
Remarkably, the seesaw model also provides an attractive explanation (which is known as the leptogenesis mechanism \cite{leptogenesis, Lreview}) for the baryon-antibaryon asymmetry of the Universe \cite{planck}
\begin{eqnarray}
Y^{}_{\rm B} \equiv \frac{n^{}_{\rm B}-n^{}_{\rm \bar B}}{s} \simeq (8.69 \pm 0.04) \times 10^{-11}  \;,
\label{2}
\end{eqnarray}
where $n^{}_{\rm B}$ ($n^{}_{\rm \bar B}$) denotes the baryon (antibaryon) number density and $s$ the entropy density. The leptogenesis mechanism works in a way as follows: a lepton-antilepton asymmetry is firstly generated from the out-of-equilibrium and CP-violating decays of right-handed neutrinos and then partly converted into the baryon-antibaryon asymmetry via the sphaleron processes.

In the basis where the charged-lepton mass matrix $M^{}_l= {\rm diag}(m^{}_e, m^{}_\mu, m^{}_\tau)$ is diagonal, the neutrino mixing matrix $U$ arises as the unitary matrix for diagonalizing $M^{}_\nu$:
\begin{eqnarray}
U^\dagger M^{}_\nu U^* = D^{}_\nu= {\rm diag}(m^{}_1, m^{}_2, m^{}_3) \;,
\label{3}
\end{eqnarray}
with $m^{}_i$ ($i=1, 2, 3$) being three light neutrino masses. Under the standard parametrization, $U$ is expressed in terms of three mixing angles $\theta^{}_{ij}$ (for $ij=12, 13, 23$), one Dirac CP phase $\delta$ and two Majorana CP phases $\rho$ and $\sigma$ as
\begin{eqnarray}
U  = \left( \begin{matrix}
c^{}_{12} c^{}_{13} & s^{}_{12} c^{}_{13} & s^{}_{13} e^{-{\rm i} \delta} \cr
-s^{}_{12} c^{}_{23} - c^{}_{12} s^{}_{23} s^{}_{13} e^{{\rm i} \delta}
& c^{}_{12} c^{}_{23} - s^{}_{12} s^{}_{23} s^{}_{13} e^{{\rm i} \delta}  & s^{}_{23} c^{}_{13} \cr
s^{}_{12} s^{}_{23} - c^{}_{12} c^{}_{23} s^{}_{13} e^{{\rm i} \delta}
& -c^{}_{12} s^{}_{23} - s^{}_{12} c^{}_{23} s^{}_{13} e^{{\rm i} \delta} & c^{}_{23}c^{}_{13}
\end{matrix} \right) \left( \begin{matrix}
e^{{\rm i}\rho} &  & \cr
& e^{{\rm i}\sigma}  & \cr
&  & 1
\end{matrix} \right) \;,
\label{4}
\end{eqnarray}
where the abbreviations $c^{}_{ij} = \cos \theta^{}_{ij}$ and $s^{}_{ij} = \sin \theta^{}_{ij}$ have been employed. Among the neutrino mass and mixing parameters,
the neutrino oscillation experiments are sensitive to three neutrino mixing angles, the neutrino mass squared differences $\Delta m^2_{ij} \equiv m^2_i - m^2_j$, and the Dirac CP phase $\delta$. Several research groups have performed global analyses of the neutrino oscillation data to extract the values of these parameters \cite{global,global2}. For definiteness, we will use the results in Ref.~\cite{global} (shown in Table~1 here) as reference values in the following numerical calculations. Note that the sign of $\Delta m^2_{31}$ remains undetermined, thereby allowing for two possible neutrino mass orderings: the normal ordering (NO) $m^{}_1 < m^{}_2 < m^{}_3$ and inverted ordering (IO) $m^{}_3 < m^{}_1 < m^{}_2$. In contrast, the neutrino oscillation experiments are completely insensitive to the absolute values of the neutrino masses and Majorana CP phases. Their values can only be inferred from certain non-oscillatory experiments such as the neutrinoless double beta decay experiments \cite{0nbb}. Unfortunately, so far there has been neither any lower constraint on the value of the lightest neutrino mass, nor any constraint on the Majorana CP phases.

Inspired by the special values of the neutrino mixing angles (e.g., the closeness of $\theta^{}_{23}$ to $\pi/4$) and a preliminary experimental hint for $\delta \sim 3\pi/2$ \cite{T2K}, in the literature the possibility that there may exist a certain flavor symmetry in the lepton sector has been attracting a lot of attention \cite{FS}. The flavor symmetries will serve as a useful guiding principle to help us organize the flavor structure of the neutrino mass model and give some interesting phenomenological consequences.
One popular candidate of them is the $\mu$-$\tau$ reflection symmetry \cite{mu-tauR, mutau2}. Under this symmetry, the neutrino mass matrices keep invariant with respect to the following transformations of three left-handed neutrino fields
\begin{eqnarray}
\nu^{}_{e} \leftrightarrow \nu^{c}_e \;, \hspace{1cm} \nu^{}_{\mu} \leftrightarrow \nu^{c}_{\tau} \;,
\hspace{1cm} \nu^{}_{\tau} \leftrightarrow \nu^{c}_{\mu} \;.
\label{5}
\end{eqnarray}
Such a symmetry leads to the following interesting predictions for the neutrino mixing parameters
\begin{eqnarray}
\theta^{}_{23} = \frac{\pi}{4} \;, \hspace{1cm} \delta = \frac{\pi}{2}  \ {\rm or} \ \frac{3\pi}{2} \;,
\hspace{1cm} \rho = 0 \ {\rm or} \ \frac{\pi}{2} \;, \hspace{1cm} \sigma = 0 \ {\rm or} \ \frac{\pi}{2} \;.
\label{6}
\end{eqnarray}
In the following numerical calculations, we will choose $\delta  = 3\pi/2$ which is more experimentally favored unless specified.

In this paper, in the commonly adopted basis of the charged lepton mass matrix $M^{}_l$ being diagonal, we will study the implications of the $\mu$-$\tau$ reflection symmetry for leptogenesis
in the type-I seesaw model with the Dirac neutrino mass matrix being diagonal \footnote{We note that Ref.~\cite{MT} has performed a similar study for the $\mu$-$\tau$ interchange symmetry.}. In this scenario, the $\mu$-$\tau$ reflection symmetry of left-handed neutrino fields is extended to the right-handed neutrino fields so that the neutrino mass matrices also keep invariant with respect to the following transformations of three right-handed neutrino fields
\begin{eqnarray}
N^{}_{e} \leftrightarrow N^{c}_e \;, \hspace{1cm} N^{}_\mu \leftrightarrow N^{c}_\tau \;,
\hspace{1cm} N^{}_\tau \leftrightarrow N^{c}_\mu \;.
\label{7}
\end{eqnarray}
In this way the Dirac neutrino mass matrix and right-handed neutrino mass matrix appear as
\begin{eqnarray}
M^{0}_{\rm D} = M^{}_0 \left( \begin{matrix}
1 & 0 & 0 \cr
0 & r & 0 \cr
 0 & 0 & r^*
\end{matrix} \right) \;, \hspace{1cm}
M^{0}_{\rm R} = \left( \begin{matrix}
A & B & B^* \cr
B & C & D \cr
B^* & D & C^*
\end{matrix} \right) \;,
\label{8}
\end{eqnarray}
with $M^{}_0$, $A$ and $D$ being real. Note that the phase of $r$ is of no physical meaning, since it can always be absorbed by the following rephasings of three left-handed neutrino fields
\begin{eqnarray}
\nu^{}_{e} \to \nu^{}_{e} \;,
\hspace{1cm} \nu^{}_{\mu} \to e^{{\rm i} {\rm arg}(r)} \nu^{}_{\mu} \;,
\hspace{1cm} \nu^{}_{\tau} \to e^{-{\rm i} {\rm arg}(r)} \nu^{}_{\tau} \;,
\label{9}
\end{eqnarray}
which are compatible with the $\mu$-$\tau$ reflection symmetry. Hence in the following we will simply take $r$ to be real and positive without loss of any generality.

Before proceeding, we would like to make the following four clarifications. Firstly, we would like to clarify our motivations for considering the scenario of $M^{}_{\rm D}$ being diagonal from the following three aspects: (1) This scenario has the advantage that it contains fewer parameters than the scenario of $M^{}_{\rm R}$ being diagonal and consequently is more predictive. To be specific, under the $\mu$-$\tau$ reflection symmetry, the neutrino mass matrices ($M^{}_{\rm D}$ and $M^{}_{\rm R}$) totally contain 8 independent real parameters in the former scenario, but 11 ones in the latter scenario. Consequently, the right-handed neutrino masses are constrained and can be predicted (see section~2 for more details) in the former scenario, while being free parameters in the latter scenario. (2) In the former scenario, both $M^{}_{\rm D}$ and $M^{}_l$ are diagonal, while the neutrino mixing comes from the non-diagonal $M^{}_{\rm R}$. In the latter scenario, both $M^{}_{\rm R}$ and $M^{}_l$ are diagonal, while the neutrino mixing comes from the non-diagonal $M^{}_{\rm D}$. It seems that the former scenario is more natural in the following aspect: $M^{}_{\rm D}$ is a Dirac-type mass matrix, so it maybe behave in a way similar to $M^{}_l$ (which is also a Dirac-type mass matrix); $M^{}_{\rm R}$ is a Majorana-type mass matrix, so it maybe behave in a way different from $M^{}_l$. (3) A diagonal $M^{}_{\rm D}$ can be naturally realized from Abelian flavor symmetries. For example, taking ${\rm Z}^{}_8$ as the Abelian flavor symmetry, if the ${\rm Z}^{}_8$-charges of three lepton doublets $L^{}_\alpha$, three right-handed charged leptons $\alpha$ and three right-handed neutrinos $N^{}_\alpha$ are as follows:
\begin{eqnarray}
L^{}_e, e, N^{}_e \sim -1 \;, \hspace{1cm} L^{}_\mu, \mu, N^{}_\mu \sim w^{}_8 \;,  \hspace{1cm} L^{}_\tau, \tau, N^{}_\tau \sim w^{3}_8 \;,
\label{10}
\end{eqnarray}
with $w^{}_8 = \exp{({\rm i} 2\pi/8)}$, then $M^{}_{\rm D}$ and $M^{}_l$ are immediately constrained to have diagonal forms (see section~2 of Ref.~\cite{MTR} for more details).

Secondly, we would like to clarify the possible values of $r$ that we will consider. Since the  $\mu$-$\tau$ reflection symmetry itself does not give any constraint on the value of $r$, in principle it can take any phenomenologically viable values: it can either be around 1, or a little larger (or smaller) than 1, or even much larger (or much smaller) than 1. In the following numerical calculations, for definiteness, we will take $r =3$ (or 0.3) as a benchmark value for the regimes that there is a mild hierarchy between $r$ and 1, and $r =10$ (or 0.1) as a benchmark value for the regimes that there is a large hierarchy between $r$ and 1. However, only when some additional flavor symmetry is incorporated can the particular case of $r=1$ be naturally realized. In fact, the case of $r=1$ can be naturally realized when the $\mu$-$\tau$ reflection symmetry is further combined with a non-Abelian discrete flavor symmetry. For example, Ref.~\cite{r1} shows that a combination of the $\mu$-$\tau$ reflection symmetry with the ${\rm S}^{}_4$ flavor symmetry can render $M^{}_{\rm D}$ to be proportional to the unity matrix (i.e., the particular case of $r=1$)
\begin{eqnarray}
M^{}_{\rm D} = M^{}_0 \left( \begin{matrix}
1 & 0 & 0 \cr
0 & 1 & 0 \cr
 0 & 0 & 1
\end{matrix} \right) \;,
\label{11}
\end{eqnarray}
and $M^{}_{\rm R}$ to have the following form:
\begin{eqnarray}
M^{}_{\rm R} = a \left( \begin{matrix}
2 & -1 & -1 \cr
-1 & 2 & -1 \cr
-1 & -1 & 2
\end{matrix} \right) +
b \left( \begin{matrix}
0 & 1 & 1 \cr
1 & 1 & 0 \cr
1 & 0 & 1
\end{matrix} \right) +
c \left( \begin{matrix}
1 & 0 & 0 \cr
0 & 0 & 1 \cr
0 & 1 & 0
\end{matrix} \right) +
d \left( \begin{matrix}
0 & -1 & 1 \cr
-1 & -2 & 0 \cr
1 & 0 & 2
\end{matrix} \right) \;,
\label{12}
\end{eqnarray}
where $a$, $b$ and $c$ are real parameters, while $d$ is purely imaginary. One can easily verify that $M^{}_{\rm R}$ in Eq.~(\ref{12}) does obey the requirements of the $\mu$-$\tau$ reflection symmetry as shown in Eq.~(\ref{8}). Not only that, it contains fewer parameters due to the presence of an additional flavor symmetry on the basis of the $\mu$-$\tau$ reflection symmetry. Correspondingly, this form of $M^{}_{\rm R}$ leads to a neutrino mixing pattern that not only has the features of the $\mu$-$\tau$ reflection symmetry but also has the features of the TM1 mixing \cite{TM}.

Thirdly, we would like to clarify our motivations for studying the consequences of the considered scenario for leptogenesis. Generally speaking, the flavor symmetries are dedicated to addressing the flavor issues of neutrinos (i.e., neutrino mixing pattern, neutrino CP violation, neutrino mass spectrum). Nevertheless, it is of important meanings to study their consequences for leptogenesis. On the one hand, if the required baryon asymmetry can be successfully reproduced from the seesaw models furnished with the studied flavor symmetries, then the requirement of leptogenesis being successful will give further constraints on the model parameters. On the other hand, if not, then one needs to invoke some other mechanisms to generate the required baryon asymmetry. Specifically for the scenario considered in this paper, leptogenesis can work successfully in the two-flavor regime and for $r \neq 1$, in which case the requirement of the required baryon asymmetry being successfully reproduced gives further constraints on the model parameters (see section~3 for more details). However, leptogenesis cannot work in the particular scenario of $r=1$ and in the scenario that the right-handed neutrino masses are below $10^9$ GeV. For these two scenarios, we will study the possibilities of leptogenesis being induced by the renormalization group evolution effects (see sections~4 and 5 for more details).

Finally, we would like to point out that Ref.~\cite{MTR} has also considered the seesaw model with neutrino mass matrices being of the forms in Eq.~(\ref{8}). But our analysis will make the following three significant improvements: (1) The study in Ref.~\cite{MTR} is concentrated on the special cases of $A=0$ or $D=0$ in Eq.~(\ref{8}). In our study, in order not to lose generality, we will relax $A$ and $D$ to be free parameters, given that the $\mu$-$\tau$ reflection symmetry itself has no constraints on them except that they have to be real parameters. (2) We will study leptogenesis in the particular scenario of $r=1$ and consider a concrete flavor-symmetry model for realizing this scenario [i.e., that shown in Eqs.~(\ref{11}-\ref{12})]. (3) We will study resonant leptogenesis at low energies which have the potential to be directly accessed by running or upcoming experiments.

The remaining part of this paper is organized as follows. In the next section, we first calculate the phenomenologically allowed values of the model parameters in Eq.~(\ref{8}) by taking account of the experimental results for the neutrino mass and mixing parameters, and then study their predictions for the right-handed neutrino masses. In section~3, we then study the consequences of the model for leptogenesis. In sections~4 and 5, we investigate the possibilities of leptogenesis being induced by the renormalization group evolution effects for the above mentioned two particular scenarios.
Finally, the summary of our main results will be given in section~6.

\begin{table}\centering
  \begin{footnotesize}
    \begin{tabular}{c|cc|cc}
     \hline\hline
      & \multicolumn{2}{c|}{Normal Ordering}
      & \multicolumn{2}{c}{Inverted Ordering }
      \\
      \cline{2-5}
      & bf $\pm 1\sigma$ & $3\sigma$ range
      & bf $\pm 1\sigma$ & $3\sigma$ range
      \\
      \cline{1-5}
      \rule{0pt}{4mm}\ignorespaces
       $\sin^2\theta^{}_{12}$
      & $0.303_{-0.012}^{+0.012}$ & $0.270 \to 0.341$
      & $0.303_{-0.012}^{+0.012}$ & $0.270 \to 0.341$
      \\[1mm]
       $\sin^2\theta^{}_{23}$
      & $0.451_{-0.016}^{+0.019}$ & $0.408 \to 0.603$
      & $0.569_{-0.021}^{+0.016}$ & $0.412 \to 0.613$
      \\[1mm]
       $\sin^2\theta^{}_{13}$
      & $0.02225_{-0.00059}^{+0.00056}$ & $0.02052 \to 0.02398$
      & $0.02223_{-0.00058}^{+0.00058}$ & $0.02048 \to 0.02416$
      \\[1mm]
       $\delta$
      & $(1.29_{-0.14}^{+0.20})\pi$ & $0.80 \pi \to 1.94 \pi$
      & $(1.53_{-0.16}^{+0.12})\pi$ & $1.08 \pi \to 1.91 \pi$
      \\[3mm]
       $\Delta m^2_{21}/(10^{-5}~{\rm eV}^2)$
      & $7.41_{-0.20}^{+0.21}$ & $6.82 \to 8.03$
      & $7.41_{-0.20}^{+0.21}$ & $6.82 \to 8.03$
      \\[3mm]
       $|\Delta m^2_{31}|/(10^{-3}~{\rm eV}^2)$
      & $2.507_{-0.027}^{+0.026}$ & $2.427 \to 2.590$
      & $2.412_{-0.025}^{+0.028}$ & $2.332 \to 2.496$
      \\[2mm]
      \hline\hline
    \end{tabular}
  \end{footnotesize}
  \caption{The best-fit values, 1$\sigma$ errors and 3$\sigma$ ranges of six neutrino
oscillation parameters extracted from a global analysis of the existing
neutrino oscillation data \cite{global}. }
\end{table}

\section{The values of model parameters and their predictions for the right-handed neutrino masses}

In this section, we first calculate the phenomenologically allowed values of the model parameters in Eq.~(\ref{8}) by taking account of the experimental results for the neutrino mass and mixing parameters, and then study their predictions for the right-handed neutrino masses.

In order to calculate the phenomenologically allowed values of the model parameters in Eq.~(\ref{8}), one can reconstruct  $M^{0}_{\rm R}$ as [see Eq.~(\ref{1})]
\begin{eqnarray}
M^{0}_{\rm R} = - M^{0T}_{\rm D} M^{-1}_{\nu} M^{0}_{\rm D} \;.
\label{2.1}
\end{eqnarray}
On the other hand, with the help of Eq.~(\ref{3}), $M^{}_\nu$ can be reconstructed in terms of the neutrino mass and mixing parameters as
\begin{eqnarray}
M^{}_\nu = U D^{}_\nu U^T \;.
\label{2.2}
\end{eqnarray}
A combination of Eqs.~(\ref{2.1}, \ref{2.2}) yields
\begin{eqnarray}
&& A  = - \frac{M^2_0}{m^{}_1} U^{*2}_{e1} - \frac{M^2_0}{m^{}_2} U^{*2}_{e2} - \frac{M^2_0}{m^{}_3} U^{*2}_{e3}  \; , \nonumber \\
&& B  = - r \frac{M^2_0}{m^{}_1} U^{*}_{e1}  U^{*}_{\mu1} - r \frac{M^2_0}{m^{}_2}  U^{*}_{e2}  U^{*}_{\mu2}  - r \frac{M^2_0}{m^{}_3}  U^{*}_{e3}  U^{*}_{\mu3}   \; , \nonumber \\
&& C  = - r^2 \frac{M^2_0}{m^{}_1} U^{*2}_{\mu1} - r^2 \frac{M^2_0}{m^{}_2} U^{*2}_{\mu 2} - r^2 \frac{M^2_0}{m^{}_3} U^{*2}_{\mu 3}  \; , \nonumber \\
&& D  = -r^2 \frac{M^2_0}{m^{}_1} U^{*}_{\mu1}  U^{*}_{\tau1} - r^2 \frac{M^2_0}{m^{}_2}  U^{*}_{\mu 2}  U^{*}_{\tau 2}  - r^2 \frac{M^2_0}{m^{}_3}  U^{*}_{\mu 3}  U^{*}_{\tau 3}   \; .
\label{2.3}
\end{eqnarray}
Taking account of the predictions of the $\mu$-$\tau$ reflection symmetry for the neutrino mixing parameters as given by Eq.~(\ref{6}), Eq.~(\ref{2.3}) becomes
\begin{eqnarray}
&& A  = - \frac{M^2_0}{m^{}_1} \eta^{}_\rho c^2_{12} c^2_{13} - \frac{M^2_0}{m^{}_2} \eta^{}_\sigma s^2_{12} c^2_{13} + \frac{M^2_0}{m^{}_3} s^{2}_{13}  \; , \nonumber \\
&& B  = \frac{r}{\sqrt 2}  \frac{M^2_0}{m^{}_1} \eta^{}_\rho c^{}_{12} c^{}_{13} \left(s^{}_{12} - {\rm i} \eta^{}_\delta c^{}_{12} s^{}_{13} \right) - \frac{r}{\sqrt 2} \frac{M^2_0}{m^{}_2} \eta^{}_\sigma s^{}_{12} c^{}_{13} \left( c^{}_{12} + {\rm i} \eta^{}_\delta s^{}_{12} s^{}_{13} \right) - \frac{r}{\sqrt 2} \frac{M^2_0}{m^{}_3} {\rm i} \eta^{}_\delta c^{}_{13} s^{}_{13}   \; , \nonumber \\
&& C  = - \frac{r^2}{2} \frac{M^2_0}{m^{}_1} \eta^{}_\rho \left(s^{}_{12}
- {\rm i} \eta^{}_\delta c^{}_{12} s^{}_{13} \right)^2 -  \frac{r^2}{2} \frac{M^2_0}{m^{}_2} \eta^{}_\sigma \left(c^{}_{12}  + {\rm i} \eta^{}_\delta s^{}_{12}
s^{}_{13} \right)^2 - \frac{r^2}{2} \frac{M^2_0}{m^{}_3} c^2_{13} \; , \nonumber \\
&& D  = - \frac{r^2}{2}  \frac{M^2_0}{m^{}_1}  \eta^{}_\rho \left( s^{2}_{12} + c^{2}_{12} s^{2}_{13}  \right) - \frac{r^2}{2} \frac{M^2_0}{m^{}_2}  \eta^{}_\sigma \left(c^{2}_{12} + s^{2}_{12} s^{2}_{13} \right) + \frac{r^2}{2}  \frac{M^2_0}{m^{}_3}  c^2_{13}   \; .
\label{2.4}
\end{eqnarray}
with $\eta^{}_\rho =1$ or $-1$ for $\rho=0$ or $\pi/2$ (and similarly for $\eta^{}_\sigma$) and $\eta^{}_\delta  = 1$ or $-1$ for $\delta = \pi/2$ or $3\pi/2$.

%%%%%%%%%%%%%%%%%%%%%% FIG 1%%%%%%%%%%%%%%%%%%%%%%
\begin{figure*}
\centering
\includegraphics[width=6.5in]{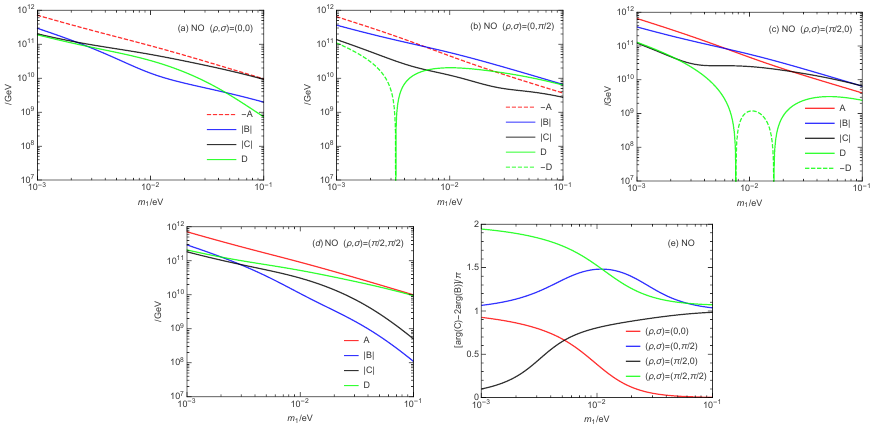}
\caption{ For the NO case, the phenomenologically allowed values of $A$, $|B|$, $|C|$, $D$ and ${\rm arg}(C)-2{\rm arg}(B)$ as functions of the lightest neutrino mass $m^{}_1$ in the cases of $(\rho, \sigma) = (0, 0), (0, \pi/2), (\pi/2, 0)$ and $(\pi/2, \pi/2)$. These results are obtained by taking $r=1$ and $M^{}_0 =1$ GeV as typical inputs. }
\label{fig1}
\end{figure*}
%%%%%%%%%%%%%%%%%%%%%%%%%%%%%%%%%%%%%%%%%%%%%%%%%%

From Eq.~(\ref{2.4}) we see that there are only three unknown parameters (i.e., the lightest neutrino mass $m^{}_{1}$ or $m^{}_3$ in the NO or IO case, $M^{}_0$ and $r$) to determine the values of $A$, $B$, $C$ and $D$, after taking account of the experimental results for the neutrino mass and mixing parameters. Accordingly, we will present the results for $A$, $B$, $C$ and $D$ as functions of $m^{}_{1}$ or $m^{}_3$ with some benchmark values of $r$ and $M^{}_0$. For the NO case, Figure~1 shows the phenomenologically allowed values of $A$, $|B|$, $|C|$ and $D$ as functions of the lightest neutrino mass $m^{}_1$ in the  cases of $(\rho, \sigma) = (0, 0), (0, \pi/2), (\pi/2, 0)$ and $(\pi/2, \pi/2)$. Note that for negative values of $A$ and $D$ we have instead shown the results for $-A$ and $-D$ and used dashed lines to denote them.
For the phases of $B$ and $C$, we have only shown the results for their particular combination ${\rm arg}(C)-2{\rm arg}(B)$, considering that it is it that keeps invariant with respect to the following rephasings of three right-handed neutrino fields (which are compatible with the $\mu$-$\tau$ reflection symmetry)
\begin{eqnarray}
N^{}_e \to  N^{}_e \;,
\hspace{1cm} N^{}_\mu \to e^{{\rm i} \varphi} N^{}_\mu \;,
\hspace{1cm} N^{}_\tau \to e^{-{\rm i} \varphi} N^{}_\tau \;,
\label{2.5}
\end{eqnarray}
and is thus of physical meaning. In obtaining the above results, we have taken $r=1$ and $M^{}_0=1$ GeV as typical inputs. The results for other values of $r$ can be obtained by noting that $|B|$ ($|C|$ and $D$) is proportional to $r$ ($r^2$) while $A$ and ${\rm arg}(C)-2{\rm arg}(B)$ are independent of $r$: the values of $|B|$ ($|C|$ and $D$) get enhanced or suppressed proportionally (quadratically) with $r$, while the values of $A$ and ${\rm arg}(C)-2{\rm arg}(B)$ keep invariant.
And the results for other values of $M^{}_0$ can be obtained by noting that $M^{0}_{\rm R}$ is simply proportional to $M^2_0$: the values of $A$, $|B|$, $|C|$ and $D$ get enhanced or suppressed quadratically with $M^{}_0$, while the values of ${\rm arg}(C)-2{\rm arg}(B)$ keep invariant.

From Figure~1 we see that there may exist zero or equal entries (which are usually taken as smoking guns for some underlying flavor physics in the literature) in $M^{0}_{\rm R}$. To be specific, $D$ will become vanishing at $m^{}_1 \simeq 0.003 $ eV in the case of $(\rho, \sigma) = (0, \pi/2)$, and at $m^{}_1 \simeq 0.007 $ eV and 0.02 eV in the case of $(\rho, \sigma) =(\pi/2, 0)$. This agrees with the finding in Ref.~\cite{MTR}. It is worth pointing out that, although this conclusion is reached in the particular case of $r=1$, it holds independently of the concrete values of $r$.
As an example for equal entries, $|B|=|C|$ holds at $m^{}_1 \simeq 0.002$ eV in the case of $(\rho, \sigma) = (0, 0)$. It is useful to note that, with the help of the rephasings of three right-handed neutrino fields in Eq.~(\ref{2.5}) by taking $\varphi ={\rm arg}(B) - {\rm arg}(C)$, one can achieve $B=C$ from $|B|=|C|$. As another example for equal entries, $|C| =D$ holds at $m^{}_1 \simeq 0.006$ eV in the case of $(\rho, \sigma) = (0, \pi/2)$, from which one can achieve $C=D$ with the help of the rephasings of three right-handed neutrino fields in Eq.~(\ref{2.5}) by taking $\varphi = - {\rm arg}(C)/2$.
However, it should be noted that the viabilities of equalities among $A$, $B$ and $C$ ($D$) rely on the concrete values of $r$, due to their different dependence behaviours on $r$. In contrast, the equality between $C$ and $D$ does not rely on the concrete values of $r$, due to their same dependence behaviours on $r$. Finally, we note that ${\rm arg}(C)-2{\rm arg}(B)$ can take the special value of $\pi/2$ or $3\pi/2$ for some specific values of $m^{}_1$, in which case one can rephase $B$ and
$C$ to be real and purely imaginary, respectively.

%%%%%%%%%%%%%%%%%%%%%% FIG 1%%%%%%%%%%%%%%%%%%%%%%
\begin{figure*}
\centering
\includegraphics[width=6.5in]{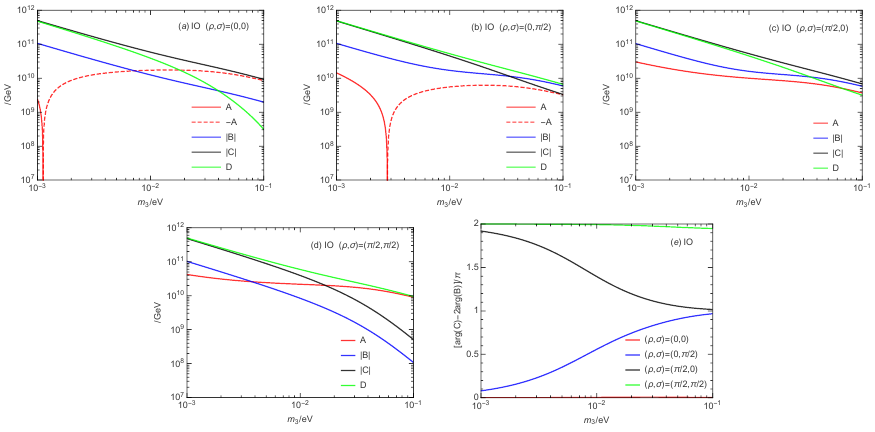}
\caption{ Same as Figure~1, except that the results here are for the IO case.  }
\label{fig2}
\end{figure*}
%%%%%%%%%%%%%%%%%%%%%%%%%%%%%%%%%%%%%%%%%%%%%%%%%%

For the IO case, Figure~2 shows the phenomenologically allowed values of $A$, $|B|$, $|C|$, $D$ and ${\rm arg}(C)-2{\rm arg}(B)$ as functions of the lightest neutrino mass $m^{}_3$ in the cases of $(\rho, \sigma) = (0, 0)$, $(0, \pi/2)$, $(\pi/2, 0)$ and $(\pi/2, \pi/2)$. In this case, it is found that $A$ will become vanishing at $m^{}_3 \simeq 0.001 $ eV in the case of $(\rho, \sigma) = (0, 0)$, and at $m^{}_3 \simeq 0.003$ eV in the case of $(\rho, \sigma) = (0, \pi/2)$. However, $C$ has no chance to be equal to $D$, independently of the concrete values of $r$.

Then, in order to obtain the right-handed neutrino masses and subsequently facilitate the leptogenesis calculations, we transform the right-handed neutrino mass matrix to a diagonal form via the unitary transformation $V$:
\begin{eqnarray}
V^T M^{0}_{\rm R} V = D^{}_N = {\rm diag}(M^{}_1, M^{}_2, M^{}_3) \;,
\label{2.6}
\end{eqnarray}
with $M^{}_I$ being three right-handed neutrino masses. In the meantime, the Dirac neutrino mass matrix is transformed to the following form
\begin{eqnarray}
M^{}_{\rm D} = M^{0}_{\rm D} V \;.
\label{2.7}
\end{eqnarray}
Given the special form of $M^{0}_{\rm R}$ in Eq.~(\ref{8}), $V$ is constrained into a form as
\begin{eqnarray}
V = \frac{1}{\sqrt{2}} \left( \begin{matrix}
1 &  & \cr
& e^{{\rm i}\phi }  & \cr
&  & e^{-{\rm i}\phi }
\end{matrix} \right)
\left( \begin{matrix}
\sqrt{2} c^{}_{x} c^{}_{y} & \sqrt{2} s^{}_{x} c^{}_{y}  & - {\rm i} \sqrt{2} \eta  s^{}_{y} \cr
-s^{}_{x} - {\rm i} \eta c^{}_{x} s^{}_{y}  & c^{}_{x} - {\rm i} \eta s^{}_{x} s^{}_{y}  &  c^{}_{y} \cr
-s^{}_{x} + {\rm i}  \eta c^{}_{x} s^{}_{y} & c^{}_{x} + {\rm i}  \eta s^{}_{x} s^{}_{y} & -c^{}_{y}
\end{matrix} \right)
\left( \begin{matrix}
e^{{\rm i}\phi^{}_1 } &  & \cr
& e^{{\rm i}\phi^{}_2 }  & \cr
&  & e^{{\rm i}\phi^{}_3 }
\end{matrix} \right) \;,
\label{2.8}
\end{eqnarray}
with $c^{}_x= \cos \theta^{}_x$, $s^{}_x= \sin \theta^{}_x$ (and similarly for $c^{}_y$ and $s^{}_y$) and $\eta = \pm 1$.
Here $\theta^{}_x$, $\theta^{}_y$ and $\phi$ are determined by
\begin{eqnarray}
&& \tan \theta^{}_{y}  = - \eta \frac{{\rm Im}[C e^{2{\rm i}\phi }] }{\sqrt{2} {\rm Re}[B e^{{\rm i}\phi } ] } \;, \hspace{1cm}  \tan 2 \theta^{}_{y}   =  - \eta \frac{ 2\sqrt{2} {\rm Im}[B e^{{\rm i}\phi } ]  }{ A + {\rm Re}[C e^{2{\rm i}\phi }] - D } \;, \nonumber  \\
&&  \tan 2 \theta^{}_{x}   =  \frac{ 2 \Delta^{}_1 }{ \Delta^{}_2 - \Delta^{}_3 } \;,
\label{2.9}
\end{eqnarray}
with
\begin{eqnarray}
&& \Delta^{}_1 = \sqrt{2} c^{}_{y} {\rm Re}[B e^{{\rm i}\phi } ] - \eta^{}_\delta s^{}_{y}{\rm Im}[C e^{2{\rm i}\phi }]  \;,  \hspace{1cm}   \Delta^{}_2 = {\rm Re}[C e^{2{\rm i}\phi }] + D \;,  \nonumber  \\
&& \Delta^{}_3 =  c^2_{y} A  + s^2_{y} \{ D - {\rm Re}[C e^{2{\rm i}\phi }] \} - 2 \sqrt{2} \eta c^{}_{y} s^{}_{y} {\rm Im}[B e^{{\rm i}\phi } ] \;.
\label{2.10}
\end{eqnarray}
And the right-handed neutrino masses are obtained as
\begin{eqnarray}
&&  M^{}_1 e^{2 {\rm i}\phi^{}_1 }  = c^2_{x} \Delta^{}_3 + s^2_{x} \Delta^{}_2 - 2 c^{}_{x} s^{}_{x} \Delta^{}_1  \;, \nonumber  \\
&& M^{}_2 e^{2 {\rm i}\phi^{}_2 } = s^2_{x} \Delta^{}_3 + c^2_{x} \Delta^{}_2 + 2 c^{}_{x} s^{}_{x} \Delta^{}_1 \;, \nonumber  \\
&& M^{}_3 e^{2 {\rm i}\phi^{}_3 } =  c^2_{y} \{ {\rm Re}[C e^{2{\rm i}\phi }] - D \} - s^2_{y} A - 2 \sqrt{2} \eta c^{}_{y} s^{}_{y} {\rm Im}[B e^{{\rm i}\phi } ] \;.
\label{2.11}
\end{eqnarray}
Due to the realness of $M^{}_I e^{2 {\rm i}\phi^{}_I }$, one simply has $\phi^{}_{I}= 0$ or $\pi/2$.

In Figures~3 and 4 (for the NO and IO cases, respectively), we have shown the possible values of three right-handed neutrino masses as functions of the lightest neutrino mass ($m^{}_1$ and $m^{}_3$, respectively) for some benchmark values of $r$ in the various cases of $(\rho, \sigma)$. In obtaining these results, we have also taken $M^{}_0=1$ GeV as a typical input, while the results for other values of $M^{}_0$ get enhanced or suppressed quadratically with $M^{}_0$. One can see that in the particular case of $r=1$ the results for three right-handed neutrino masses do not rely on the values of $(\rho, \sigma)$. This is because for $r=1$ one has $V={\rm i}U$ and $M^{}_I = M^{2}_0/m^{}_i$ [see Eq.~(\ref{2.1})]. It is interesting to note that there may exist two nearly degenerate right-handed neutrinos in some parameter ranges, in which cases low-scale resonant leptogenesis may be realized (see section~5).

%%%%%%%%%%%%%%%%%%%%%% FIG 3%%%%%%%%%%%%%%%%%%%%%%
\begin{figure*}
\centering
\includegraphics[width=6.5in]{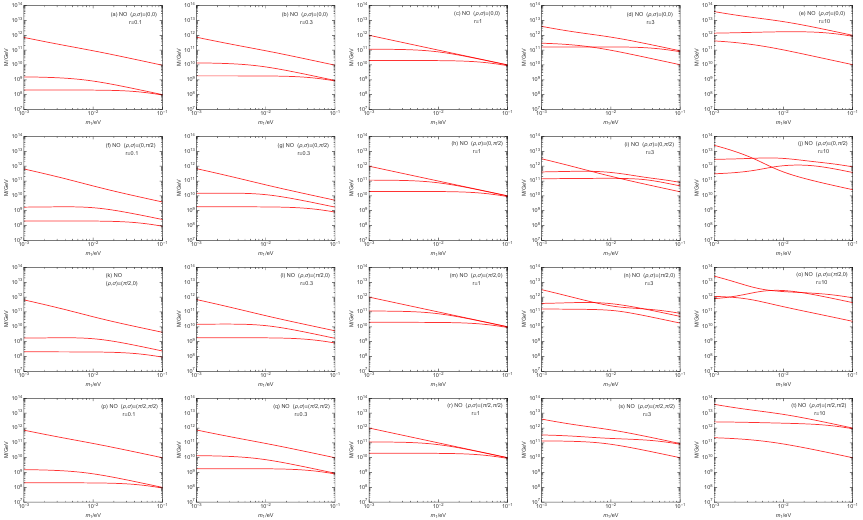}
\caption{ For the NO case, the possible values of three right-handed neutrino masses as functions of the lightest neutrino mass $m^{}_1$ for the benchmark values 0.1, 0.3, 1, 3 and 10 of $r$ in the cases of $(\rho, \sigma) = (0, 0), (0, \pi/2), (\pi/2, 0)$ and $(\pi/2, \pi/2)$. These results are obtained by taking $M^{}_0 =1$ GeV as a typical input. }
\label{fig3}
\end{figure*}
%%%%%%%%%%%%%%%%%%%%%%%%%%%%%%%%%%%%%%%%%%%%%%%%%%

%%%%%%%%%%%%%%%%%%%%%% FIG 41%%%%%%%%%%%%%%%%%%%%%%
\begin{figure*}
\centering
\includegraphics[width=6.5in]{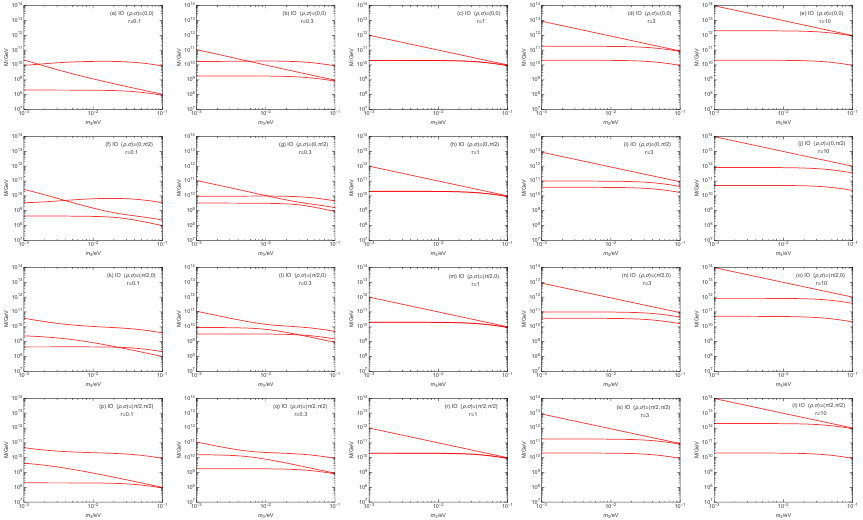}
\caption{ Same as Figure~3 except that the results here are for the IO case.  }
\label{fig4}
\end{figure*}
%%%%%%%%%%%%%%%%%%%%%%%%%%%%%%%%%%%%%%%%%%%%%%%%%%

\section{Consequences for leptogenesis}

Now, let us study the consequences of the model considered in this paper for leptogenesis.

As is known, according to the temperature ranges where leptogenesis takes place (approximately the right-handed neutrino mass scale), there are the following three distinct leptogenesis regimes  \cite{flavor}. (1) One-flavor regime: in the temperature range above $10^{12}$ GeV where the charged lepton Yukawa $y^{}_\alpha$ interactions have not yet entered thermal equilibrium, three lepton flavors are indistinguishable from one another so that they should be treated in a universal way. In this regime, the final baryon asymmetry from the right-handed neutrino $N^{}_I$ can be calculated according to
\begin{eqnarray}
Y^{}_{\rm B} = c d \varepsilon^{}_I \kappa(\widetilde m^{}_I)  \;,
\label{3.1}
\end{eqnarray}
where $c = -28/79$ describes the transition efficiency from the lepton-antilepton asymmetry to the baryon-antibaryon asymmetry via the sphaleron processes, and $d \simeq 4 \times 10^{-3}$ measures the ratio of the equilibrium number density of $N^{}_I$ to the entropy density. And $\varepsilon^{}_I$ is a sum (over three lepton flavors) of the following flavored CP asymmetries (between the decay rates of $N^{}_I \to L^{}_\alpha + H$ and their CP-conjugate processes $N^{}_I \to \overline{L}^{}_\alpha + \overline{H}$)
\begin{eqnarray}
&& \varepsilon^{}_{I \alpha} = \frac{1}{8\pi (M^\dagger_{\rm D}
M^{}_{\rm D})^{}_{II} v^2} \sum^{}_{J \neq I} \left\{ {\rm Im}\left[(M^*_{\rm D})^{}_{\alpha I} (M^{}_{\rm D})^{}_{\alpha J}
(M^\dagger_{\rm D} M^{}_{\rm D})^{}_{IJ}\right] {\cal F} \left( \frac{M^2_J}{M^2_I} \right) \right. \nonumber \\
&& \hspace{1.cm}
+ \left. {\rm Im}\left[(M^*_{\rm D})^{}_{\alpha I} (M^{}_{\rm D})^{}_{\alpha J} (M^\dagger_{\rm D} M^{}_{\rm D})^*_{IJ}\right] {\cal G}  \left( \frac{M^2_J}{M^2_I} \right) \right\} \; ,
\label{3.2}
\end{eqnarray}
with ${\cal F}(x) = \sqrt{x} \{(2-x)/(1-x)+ (1+x) \ln [x/(1+x)] \}$ and ${\cal G}(x) = 1/(1-x)$.
Finally, $\kappa(\widetilde m^{}_I)$ is the efficiency factor which is smaller than 1 due to the washout effects. Its concrete value depends on the washout mass parameter
\begin{eqnarray}
\widetilde m^{}_I = \sum^{}_\alpha \widetilde m^{}_{I \alpha} = \sum^{}_\alpha  \frac{|(M^{}_{\rm D})^{}_{\alpha I}|^2}{M^{}_I} \;,
\label{3.3}
\end{eqnarray}
and can be numerically calculated by solving the relevant Boltzmann equations \cite{Lreview}.
(2) Two-flavor regime: in the temperature range $10^{9}$---$10^{12}$ GeV where the $y^{}_\tau$-related interactions have entered thermal equilibrium, the $\tau$ flavor is distinguishable from the other two flavors which remain indistinguishable from each other so that there are effectively two flavors (i.e., the $\tau$ flavor and a coherent superposition of the $e$ and  $\mu$ flavors). In this regime, the final baryon asymmetry from $N^{}_I$ can be calculated according to
\begin{eqnarray}
Y^{}_{\rm B}
=  c d \left[ \varepsilon^{}_{I \gamma} \kappa \left(\frac{417}{589} \widetilde m^{}_{I \gamma} \right) + \varepsilon^{}_{I \tau} \kappa \left(\frac{390}{589} \widetilde m^{}_{I \tau} \right) \right]
 \;,
\label{3.4}
\end{eqnarray}
with $\varepsilon^{}_{I \gamma} = \varepsilon^{}_{I e} + \varepsilon^{}_{I \mu}$ and $\widetilde m^{}_{I \gamma} = \widetilde m^{}_{I e} + \widetilde m^{}_{I \mu}$. (3) Three-flavor regime: in the temperature range below $10^{9}$ GeV where the $y^{}_\mu$-related interactions have also entered thermal equilibrium, all the three lepton flavors are distinguishable from one another so that they should be treated separately. In this regime, the final baryon asymmetry from $N^{}_I$ can be calculated according to
\begin{eqnarray}
Y^{}_{\rm B} = c d \left[ \varepsilon^{}_{I e} \kappa \left(\frac{453}{537} \widetilde m^{}_{I e} \right) + \varepsilon^{}_{I \mu} \kappa \left(\frac{344}{537} \widetilde m^{}_{I \mu} \right) + \varepsilon^{}_{I \tau} \kappa \left(\frac{344}{537} \widetilde m^{}_{I\tau} \right) \right] \; .
\label{3.5}
\end{eqnarray}

For the model considered in this paper, in the mass basis of right-handed neutrinos $M^{}_{\rm D}$ is given by Eq.~(\ref{2.7}), which leads to
\begin{eqnarray}
\varepsilon^{}_{I} =0 \;, \hspace{1cm} \varepsilon^{}_{I e} =0 \;, \hspace{1cm} \varepsilon^{}_{I \mu} =- \varepsilon^{}_{I \tau} \;,  \hspace{1cm} \widetilde m^{}_{I \mu} = \widetilde m^{}_{I \tau} \;.
\label{3.6}
\end{eqnarray}
As a direct result, in the one-flavor and three-flavor leptogenesis regimes, one would arrive at $Y^{}_{\rm B} =0$ [see Eqs.~(\ref{3.1}, \ref{3.5})]. Namely, leptogenesis cannot work in these two regimes. Fortunately, in the two-flavor regime, a successful leptogenesis may become possible \cite{MN}: taking into account the relations in Eq.~(\ref{3.6}), Eq.~(\ref{3.4}) is simplified to
\begin{eqnarray}
Y^{}_{\rm B}
=  c d \varepsilon^{}_{I \mu} \left[ \kappa \left(\frac{417}{589} \widetilde m^{}_{I \gamma} \right)  -\kappa \left(\frac{390}{589} \widetilde m^{}_{I \tau} \right) \right]
 \;,
\label{3.7}
\end{eqnarray}
which shows that unless $390 \widetilde m^{}_{I \tau}$ coincides with $417 \widetilde m^{}_{I \gamma}$, $Y^{}_{\rm B}$ will be non-vanishing. Hence we will work in the two-flavor regime in the following discussions. For $M^{}_{\rm D}$ in Eq.~(\ref{2.7}), $\varepsilon^{}_{1 \mu}$ is explicitly given by
\begin{eqnarray}
\varepsilon^{}_{1 \mu} = \frac{\eta M^2_0 r^2 (1-r^2) c^{}_x s^{}_x c^2_y s^{}_y }{16\pi v^2 [ c^2_x c^2_y + r^2 (s^2_x + c^2_x s^2_y) ] } \left[ \eta^{}_1 \eta^{}_2 {\cal F} \left( \frac{M^2_2}{M^2_1} \right) + \eta^{}_1 \eta^{}_3 {\cal F} \left( \frac{M^2_3}{M^2_1} \right) +  {\cal G}  \left( \frac{M^2_2}{M^2_1} \right)  - {\cal G}  \left( \frac{M^2_3}{M^2_1} \right) \right] \; ,
\label{3.8}
\end{eqnarray}
with $\eta^{}_1 = 1$ or $-1$ for $\phi^{}_1 = 0$ or $\pi/2$ (and similarly for $\eta^{}_2$). It is direct to see that, in the particular case of $r=1$, $\varepsilon^{}_{1 \mu}$ (and similarly for $\varepsilon^{}_{2 \mu}$ and $\varepsilon^{}_{3 \mu}$) would be vanishing and thus leptogenesis would fail. This can be understood from the fact that in the case of $r=1$ $M^{}_{\rm D}$ would be proportional to a unitary matrix (i.e., $V$), and this would lead to the orthogonality relations among different columns of $M^{}_{\rm D}$ [i.e., $(M^\dagger_{\rm D} M^{}_{\rm D})^{}_{IJ}=0$]. For this reason, we will consider the general cases of $r\neq 1$ in the rest part of this section, but will study the possibility of leptogenesis being induced by the renormalization group evolution effects for the particular case of $r =1$ in the next section.

As we have seen from Eq.~(\ref{2.4}), there are only three unknown parameters (i.e., the lightest neutrino mass $m^{}_{1}$ or $m^{}_3$, $M^{}_0$ and $r$) to determine the values of $A$, $B$, $C$ and $D$ (after taking account of the experimental results for the neutrino mass and mixing parameters), which subsequently determine the values of three right-handed neutrino masses. Therefore, in section~2 we have presented the results for $A$, $B$, $C$, $D$ and resultant three right-handed neutrino masses as functions of $m^{}_{1}$ or $m^{}_3$ with some benchmark values of $r$ and $M^{}_0$. Now, by imposing the requirement that the observed value of $Y^{}_{\rm B}$ should be successfully reproduced, $M^{}_0$ and subsequently three right-handed neutrino masses can be determined as functions of $m^{}_{1}$ or $m^{}_3$ for given values of $r$. Accordingly, for some benchmark values of $r$, in Figure~5 we have shown the values of the lightest right-handed neutrino mass $M^{}_1$ that allow for a reproduction of the observed value of $Y^{}_{\rm B}$ as functions of the lightest neutrino mass $m^{}_1$ or $m^{}_3$ in the cases of $(\rho, \sigma) = (0, 0), (0, \pi/2), (\pi/2, 0)$ and $(\pi/2, \pi/2)$. In the NO case, it is found that only the cases of $(\rho, \sigma) = (0, 0)$ and $(\pi/2, \pi/2)$ allow for leptogenesis to be successful, while the cases of $(\rho, \sigma) = (0, \pi/2)$ and $(\pi/2, 0)$ do not allow for a successful leptogenesis. And there exists an upper bound about 0.03 eV for $m^{}_1$. These results have important implications for the neutrinoless double beta decay experiments: Figure~6(a) shows the resultant values of the effective neutrino mass which controls the rates of the neutrinoless double beta decays
\begin{eqnarray}
\left| (M^{}_{\nu})^{}_{ee} \right| = \left| m^{}_1 e^{2{\rm i}\rho} c^2_{12} c^2_{13} + m^{}_2 e^{2{\rm i}\sigma} s^2_{12} c^2_{13} + m^{}_3 s^2_{13} e^{-2{\rm i}\delta} \right| \;,
\label{3.9}
\end{eqnarray}
for these two cases of $(\rho, \sigma)$. We see that in the case of $(\rho, \sigma) = (\pi/2, \pi/2)$ the values of $\left| (M^{}_{\nu})^{}_{ee} \right|$ are more likely to be probed by the neutrinoless double beta decay experiments. This is because in this case the three terms of $\left| (M^{}_{\nu})^{}_{ee} \right|$ add constructively.
Furthermore, note that only for $r>1$ can the observed value of $Y^{}_{\rm B}$ be successfully reproduced. Of course, the allowed parameter space of $M^{}_1$ and $m^{}_1$ would gradually vanish in the limit of $r \to 1$: when $r$ is close to 1, the observed value of $Y^{}_{\rm B}$ can be successfully reproduced only for $m^{}_1 \sim 0.02$ or 0.03 eV in the case of $(\rho, \sigma) = (0, 0)$ or $(\pi/2, \pi/2)$ and $M^{}_1$ close to $10^{12}$ GeV (which is the upper boundary for the two-flavor leptogenesis regime to hold). When $r$ becomes much larger than 1, the observed value of $Y^{}_{\rm B}$ may be successfully reproduced for $m^{}_1 \lesssim 0.03$ eV or $\lesssim 0.02$ eV in the case of $(\rho, \sigma) = (0, 0)$ or $(\pi/2, \pi/2)$ and $M^{}_1 \gtrsim 10^{11}$ GeV.

In the IO case, it turns out that only the cases of $(\rho, \sigma) = (0, \pi/2)$ and $(\pi/2, 0)$ allow for leptogenesis to be successful, while the cases of $(\rho, \sigma) = (0, 0)$ and $(\pi/2, \pi/2)$ do not allow for a successful leptogenesis. And there exists an upper bound about 0.1 eV or 0.02 eV for $m^{}_3$ in the case of $(\rho, \sigma) = (0, \pi/2)$ or $(\pi/2, 0)$. Figure~6(b) shows the resultant values of $\left| (M^{}_{\nu})^{}_{ee} \right|$
for these two cases of $(\rho, \sigma)$. We see that the values of $\left| (M^{}_{\nu})^{}_{ee} \right|$ are nearly equal for these two cases of $(\rho, \sigma)$. This is because in the IO case the third term of $\left| (M^{}_{\nu})^{}_{ee} \right|$ are negligibly small due to the simultaneous suppression of $m^{}_3$ and $s^2_{13}$.
Furthermore, it is found that only for $r<1$ can the observed value of $Y^{}_{\rm B}$ be successfully reproduced. Of course, the allowed parameter space of $M^{}_1$ and $m^{}_1$ would also gradually vanish in the limit of $r \to 1$: when $r$ is close to 1, the observed value of $Y^{}_{\rm B}$ can be successfully reproduced only for $m^{}_3 \sim 0.1$ or 0.001 eV in the case of $(\rho, \sigma) = (0, \pi/2)$ or $(\pi/2, 0)$ and $M^{}_1$ close to $10^{12}$ GeV. When $r$ becomes much smaller than 1, the observed value of $Y^{}_{\rm B}$ may be successfully reproduced for $m^{}_3 \lesssim 0.05$ eV or $\lesssim 0.02$ eV and $M^{}_1 \gtrsim 10^{11}$ GeV or $\gtrsim 4 \times 10^{11}$ GeV in the case of $(\rho, \sigma) = (0, \pi/2)$ or $(\pi/2, 0)$.

%%%%%%%%%%%%%%%%%%%%%% FIG 3%%%%%%%%%%%%%%%%%%%%%%
\begin{figure*}
\centering
\includegraphics[width=6.5in]{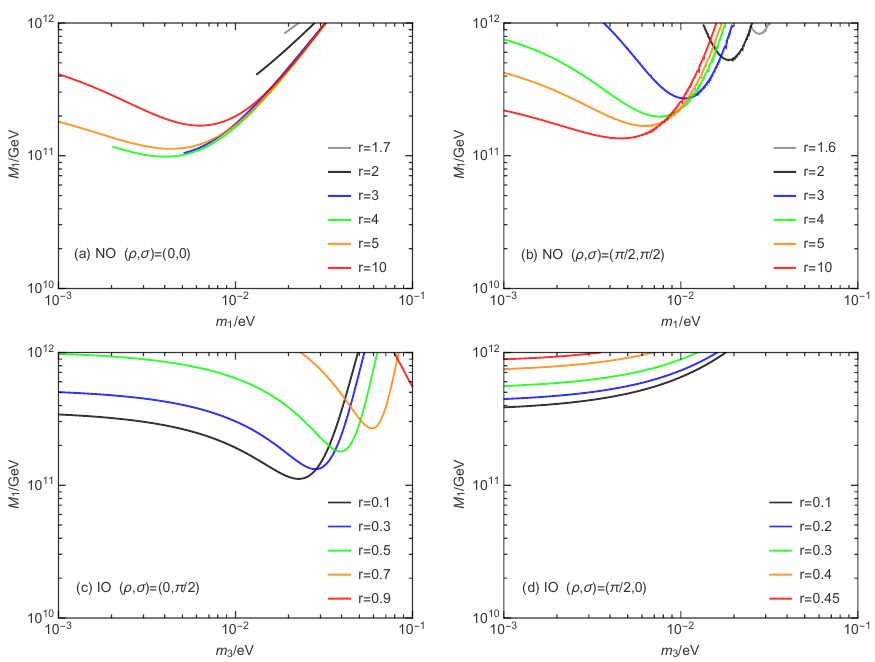}
\caption{ In the NO (IO) case, for some benchmark values of $r$, the values of the lightest right-handed neutrino mass $M^{}_1$ that allow for a reproduction of the observed value of $Y^{}_{\rm B}$ as functions of the lightest neutrino mass $m^{}_1$ ($m^{}_3$) in the cases of $(\rho, \sigma) = (0, 0), (0, \pi/2), (\pi/2, 0)$ and $(\pi/2, \pi/2)$. }
\label{fig5}
\end{figure*}
%%%%%%%%%%%%%%%%%%%%%%%%%%%%%%%%%%%%%%%%%%%%%%%%%%

%%%%%%%%%%%%%%%%%%%%%% FIG 3%%%%%%%%%%%%%%%%%%%%%%
\begin{figure*}
\centering
\includegraphics[width=6.5in]{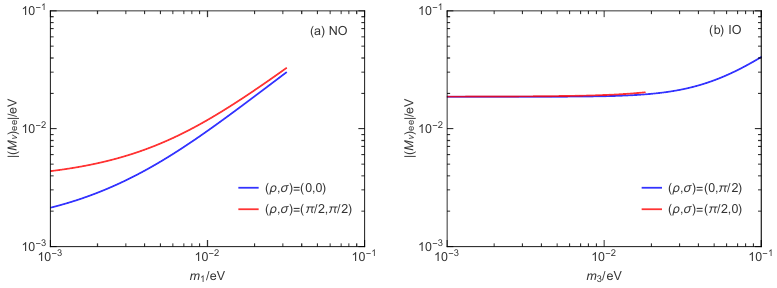}
\caption{ In the NO (IO) case, under the requirement that the observed value of $Y^{}_{\rm B}$ should be successfully reproduced, the allowed values of the effective neutrino mass which controls the rates of the neutrinoless double beta decays as functions of the lightest neutrino mass $m^{}_1$ ($m^{}_3$). }
\label{fig6}
\end{figure*}
%%%%%%%%%%%%%%%%%%%%%%%%%%%%%%%%%%%%%%%%%%%%%%%%%%

\section{Renormalization group evolution induced leptogenesis for the particular case of $r=1$}

As we have seen, for the particular case of $r=1$, leptogenesis cannot work due to the orthogonality relations among different columns of $M^{}_{\rm D}$. In this section, we study the possibility of leptogenesis being induced by the renormalization group evolution effects for this case: in the literature, the flavor symmetries are usually placed at a very high energy scale $\Lambda^{}_{\rm FS}$. When dealing with leptogenesis which takes place around the right-handed neutrino mass scale $M^{}_I$, the renormalization group evolution effects should be taken into account if there is a large gap between $\Lambda^{}_{\rm FS}$ and $M^{}_I$.

Before proceeding, we would like to clarify our motivations for considering the scenario that leptogenesis is induced by the renormalization group evolution effects from the following three aspects: (1) This effect is inevitable, provided that there is a considerable gap between the flavor-symmetry scale and the right-handed neutrino mass scale. (2) This effect is minimal in the sense that it does not need to introduce additional parameters (except for the flavor-symmetry scale). (3) Furthermore, we note that in the case of $r=1$ leptogenesis still cannot work when the corrections for the flavor symmetries arise from a non-diagonal charged lepton mass matrix. This is because in this case the Dirac neutrino mass matrix (i.e., the neutrino Yukawa matrix) would remain to be proportional to a unitary matrix after one goes back to the mass basis of the charged leptons via a unitary transformation.

At the one-loop level, the running behaviour of the Dirac neutrino mass matrix is described by \cite{ynu}
\begin{eqnarray}
16 \pi^2 \frac{d M^{}_{\rm D}}{dt} = \left[ \frac{3}{2} Y^{}_\nu Y^\dagger_\nu - \frac{3}{2} Y^{}_l Y^\dagger_l + {\rm Tr} \left( 3 Y^{}_u Y^\dagger_u + 3 Y^{}_d Y^\dagger_d +  Y^{}_\nu Y^\dagger_\nu + Y^{}_l Y^\dagger_l \right) - \frac{9}{20} g^2_1 - \frac{9}{4} g^2_2 \right]  M^{}_{\rm D} \;,
\label{4.1}
\end{eqnarray}
in the SM, and
\begin{eqnarray}
16 \pi^2 \frac{d M^{}_{\rm D}}{dt} = \left[ 3 Y^{}_\nu Y^\dagger_\nu + Y^{}_l Y^\dagger_l + {\rm Tr} \left( 3 Y^{}_u Y^\dagger_u + Y^{}_\nu Y^\dagger_\nu  \right) - \frac{3}{5} g^2_1 - 3 g^2_2 \right]  M^{}_{\rm D} \;,
\label{4.2}
\end{eqnarray}
in the Minimal Supersymmetric Standard Model (MSSM). Here $t$ denotes $\ln(\mu/\Lambda^{}_{\rm FS})$ with $\mu$ being the renormalization scale, $Y^{}_{u, d}$ are the up-type-quark and down-type-quark Yukawa matrices and $g^{}_{1, 2}$ are the gauge couplings. In the basis of $M^{}_l$ being diagonal, the Yukawa coupling matrix for three charged leptons
is given by $Y^{}_l = {\rm Diag} (y^{}_e, y^{}_\mu, y^{}_\tau)$.

In the SM framework, an integration of Eq.~(\ref{4.1}) enables us to obtain the Dirac neutrino mass matrix $M^{}_{\rm D}(M^{}_I)$ at the right-handed neutrino mass scale from its counterpart $M^{}_{\rm D}(\Lambda^{}_{\rm FS})$ at the flavor-symmetry scale as \cite{IRGE}
\begin{eqnarray}
M^{}_{\rm D} (M^{}_0) = I^{}_{0} \left( \begin{array}{ccc}
1+\Delta^{}_{e} &   &  \cr
 & 1 +\Delta^{}_{\mu} &  \cr
 &  &  1+\Delta^{}_{\tau} \cr
\end{array} \right)
M^{}_{\rm D} (\Lambda^{}_{\rm FS}) \;,
\label{4.3}
\end{eqnarray}
where
\begin{eqnarray}
&& I^{}_{0}  =  {\rm exp} \left\{ - \frac{1}{16 \pi^2} \int^{\ln (\Lambda^{}_{\rm FS}/M^{}_I)}_{0} \left[ {\rm Tr} \left( 3 Y^{}_u Y^\dagger_u + 3 Y^{}_d Y^\dagger_d +  Y^{}_\nu Y^\dagger_\nu + Y^{}_l Y^\dagger_l \right) - \frac{9}{20} g^2_1 - \frac{9}{4} g^2_2 \right] \ {\rm dt} \right\} \;, \nonumber \\
&& \Delta^{}_{\alpha}   =   \frac{3}{32 \pi^2}\int^{\ln (\Lambda^{}_{\rm FS}/M^{}_I)}_{0} y^2_{\alpha} \ {\rm dt} \simeq \frac{3}{32 \pi^2} y^2_{\alpha} \ln \left(\frac{\Lambda^{}_{\rm FS}}{M^{}_I} \right) \;.
\label{4.4}
\end{eqnarray}
We see that $I^{}_0$ is just an overall rescaling factor and it can be absorbed by the redefinitions of the model parameters and thus can be dropped without affecting the final results. In contrast, in spite of being small, $\Delta^{}_\alpha$ can modify the structure of $M^{}_{\rm D}$, inducing remarkable consequences for leptogenesis as we will see soon. Numerically, one has $\Delta^{}_{e} \ll \Delta^{}_{\mu} \ll \Delta^{}_{\tau} \ll 1$ (as a result of $y^{}_e \ll y^{}_\mu \ll y^{}_\tau \ll 1$), so it is an excellent approximation for us to only keep $\Delta^{}_{\tau}$ in the following calculations. Furthermore, for $\Lambda^{}_{\rm FS}/M^{}_I =100$ as a benchmark value, one has $\Delta^{}_\tau \sim 4.4\times 10^{-6}$ in the SM. On the other hand, in the MSSM $y^{2}_\tau = (1+ \tan^2{\beta}) m^2_\tau/v^2$ and consequently $\Delta^{}_\tau$ can be greatly enhanced by a large $\tan{\beta}$ value. To be explicit, the value of $\Delta^{}_{\tau}$ depends on the large $\tan{\beta}$ values in a way as
\begin{eqnarray}
\Delta^{}_{\tau} \sim - 0.022 \left( \displaystyle \frac{\tan{\beta}}{50} \right)^2 \;.
\label{4.5}
\end{eqnarray}
Before proceeding, we point out that in the MSSM, in spite of the doubling of the particle spectrum and of the large number of new processes involving superpartners, one does not expect major numerical changes with respect to the non-supersymmetric case. To be more specific, for given values of $M^{}_I$, $Y^{}_\nu$ and $Y^{}_l$, the total effect of supersymmetry on the final baryon number asymmetry can simply be summarized as a constant factor (for a detailed explanation, see section~10.1 of the third reference in Ref.~\cite{Lreview}):
\begin{eqnarray}
\left. \frac{Y^{\rm MSSM}_{\rm B}}{Y^{\rm SM}_{\rm B}} \right|^{}_{M^{}_I, Y^{}_\nu, Y^{}_l} \simeq \left\{ \begin{array}{l} \sqrt{2} \hspace{0.5cm} ({\rm in \ strong \ washout \ regime} ) \; ; \\ 2 \sqrt{2} \hspace{0.5cm} ({\rm in \ weak \ washout \ regime}) \; . \end{array} \right.
\label{4.5.2}
\end{eqnarray}
In the following leptogenesis calculations within the MSSM framework, we will take into account of such a factor.

For the particular case of $r=1$ under consideration, the renormalization group evolution effects will break the orthogonality relations among different columns of $M^{}_{\rm D}$ (due to the differences among $y^{}_\alpha$) and consequently induce leptogenesis to work (see also Refs.~\cite{rgeL}). Furthermore, the renormalization group evolution effects will also break the $\mu$-$\tau$ reflection symmetry (due to the difference between $y^{}_\mu$ and $y^{}_\tau$) and consequently induce leptogenesis to work even in the one-flavor and three-flavor leptogenesis regimes in which leptogenesis would otherwise not work. To be explicit, in the one-flavor leptogenesis regime, for an $M^{}_{\rm D}(\Lambda^{}_{\rm FS})$ as given by Eq.~(\ref{2.7}), $\varepsilon^{}_{1}$ becomes non-vanishing as
\begin{eqnarray}
\varepsilon^{}_{1} = \Delta^2_\tau \frac{\eta  M^2_0 c^{}_x s^{}_x c^2_y s^{}_y }{4\pi v^2  } \left[ \eta^{}_1 \eta^{}_2 {\cal F} \left( \frac{M^2_2}{M^2_1} \right) + \eta^{}_1 \eta^{}_3 {\cal F} \left( \frac{M^2_3}{M^2_1} \right)  \right] \;.
\label{4.6}
\end{eqnarray}
But it is suppressed by $\Delta^2_\tau$, so the observed value of $Y^{}_{\rm B}$ cannot be successfully reproduced. In the two-flavor leptogenesis regime, the relevant flavored CP asymmetries  $\varepsilon^{}_{1\gamma}$ and $\varepsilon^{}_{1\tau}$ are given by
\begin{eqnarray}
\varepsilon^{}_{1 \gamma} = - \varepsilon^{}_{1 \tau} = - \Delta^{}_\tau \frac{ \eta M^2_0  c^{}_x s^{}_x c^2_y s^{}_y }{8\pi v^2  } \left[ \eta^{}_1 \eta^{}_2 {\cal F} \left( \frac{M^2_2}{M^2_1} \right) + \eta^{}_1 \eta^{}_3 {\cal F} \left( \frac{M^2_3}{M^2_1} \right)  \right] \; .
\label{4.7}
\end{eqnarray}
Now the relevant CP asymmetries are only suppressed by $\Delta^{}_\tau$, so the observed value of $Y^{}_{\rm B}$ may be successfully reproduced. But in the SM framework, due to the smallness of $\Delta^{}_\tau$, the
resultant $Y^{}_{\rm B}$ cannot reproduce its observed value.
In contrast, in the MSSM framework the observed value of $Y^{}_{\rm B}$ may be  successfully reproduced due to the following two enhancement effects from a large $\tan \beta$ value: on the one hand, as shown in Eq.~(\ref{4.5}), a large $\tan\beta$ value will greatly enhance the size of $\Delta^{}_\tau$ which directly control the strengths of relevant flavored CP asymmetries; on the other hand, a large $\tan\beta$ value will lift the upper boundary for the two-flavor leptogenesis regime to hold from $10^{12}$ GeV to $(1+\tan^2 \beta) 10^{12}$ GeV \cite{MSSM}, via which the enlargement of the allowed right-handed neutrino mass scale will also enhance the sizes of relevant flavored CP asymmetries. However, it should be noted that the right-handed neutrino mass should not exceed $10^{14}$ GeV because otherwise the $\Delta L=2$ processes mediated by the right-handed neutrinos would greatly suppress the efficiency of leptogenesis \cite{Lreview}.
Similarly, in the three-flavor leptogenesis regime, the renormalization group evolution effects may also induce leptogenesis to work. In this regime, the relevant flavored CP asymmetries $\varepsilon^{}_{1e}$ and $\varepsilon^{}_{1\mu}$ are given by
\begin{eqnarray}
&& \varepsilon^{}_{1 e} = - \Delta^{}_\tau \frac{ \eta M^2_0  c^{}_x s^{}_x c^2_y s^{}_y }{8\pi v^2  } \left[ \eta^{}_1 \eta^{}_2 {\cal F} \left( \frac{M^2_2}{M^2_1} \right) + \eta^{}_1 \eta^{}_3 {\cal F} \left( \frac{M^2_3}{M^2_1} \right) -  {\cal G}  \left( \frac{M^2_2}{M^2_1} \right)  + {\cal G}  \left( \frac{M^2_3}{M^2_1} \right) \right] \; , \nonumber \\
&& \varepsilon^{}_{1 \mu} = - \Delta^{}_\tau \frac{ \eta M^2_0  c^{}_x s^{}_x c^2_y s^{}_y }{8\pi v^2  } \left[  {\cal G}  \left( \frac{M^2_2}{M^2_1} \right)  - {\cal G}  \left( \frac{M^2_3}{M^2_1} \right) \right] \; ,
\label{4.8}
\end{eqnarray}
while $\varepsilon^{}_{1\tau}$ is same as in Eq.~(\ref{4.7}). Note that in the MSSM framework the upper boundary for the three-flavor leptogenesis regime to hold will be analogously lifted from $10^{9}$ GeV to $(1+\tan^2 \beta) 10^{9}$ GeV.

%%%%%%%%%%%%%%%%%%%%%% FIG 3%%%%%%%%%%%%%%%%%%%%%%
\begin{figure*}
\centering
\includegraphics[width=6.5in]{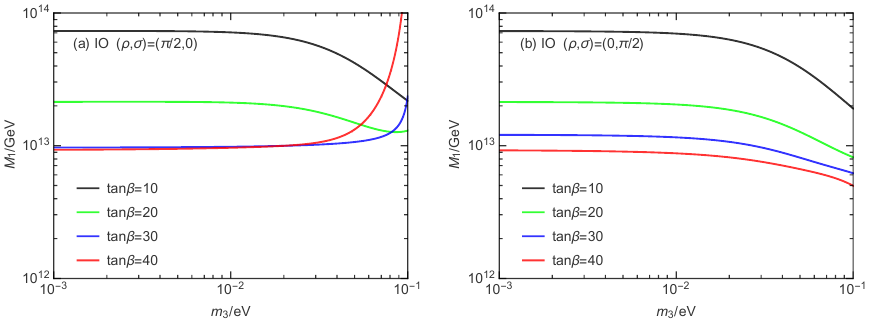}
\caption{ In the IO case, for some benchmark values of $\tan \beta$, the values of the lightest right-handed neutrino mass that allow for a reproduction of the observed value of $Y^{}_{\rm B}$ as functions of the lightest neutrino mass $m^{}_3$ in the cases of $(\rho, \sigma) =  (\pi/2, 0)$ and $(0, \pi/2)$. }
\label{fig7}
\end{figure*}
%%%%%%%%%%%%%%%%%%%%%%%%%%%%%%%%%%%%%%%%%%%%%%%%%%

For some benchmark values of $\tan \beta$, Figure~7 has shown the values of the lightest right-handed neutrino mass $M^{}_1$ that allow for a reproduction of the observed value of $Y^{}_{\rm B}$ as functions of the lightest neutrino mass in the phenomenologically allowed cases of $(\rho, \sigma)$. Note that in the MSSM values of $\tan \beta$ larger than 55 are unnatural since the bottom quark mass cannot be reproduced with an order one coupling. So we will only consider values of $\tan \beta$ smaller than 50. It is found that only in the IO case and for $(\rho, \sigma)=(0, \pi/2)$ or $(\pi/2, 0)$ can the observed value of $Y^{}_{\rm B}$ be successfully reproduced.
And the observed value of $Y^{}_{\rm B}$ can be successfully reproduced for a moderately large $\tan\beta$ ($\gtrsim 10$) and almost the whole $m^{}_3$ range.

Finally, we consider the specific flavor-symmetry model shown in Eqs.~(\ref{11}-\ref{12}) which can naturally realize the particular scenario of $r=1$. Since this model is subject to more constraints than the above generic cases featuring $r=1$, it is natural to expect that only in the IO case can it have chance to successfully reproduce the observed value of $Y^{}_{\rm B}$.
For this model, also for some benchmark values of $\tan \beta$, Figure~8 has shown the values of the lightest right-handed neutrino mass $M^{}_1$ that allow for a reproduction of the observed value of $Y^{}_{\rm B}$ as functions of the lightest neutrino mass $m^{}_3$.

%%%%%%%%%%%%%%%%%%%%%% FIG 3%%%%%%%%%%%%%%%%%%%%%%
\begin{figure*}
\centering
\includegraphics[width=5in]{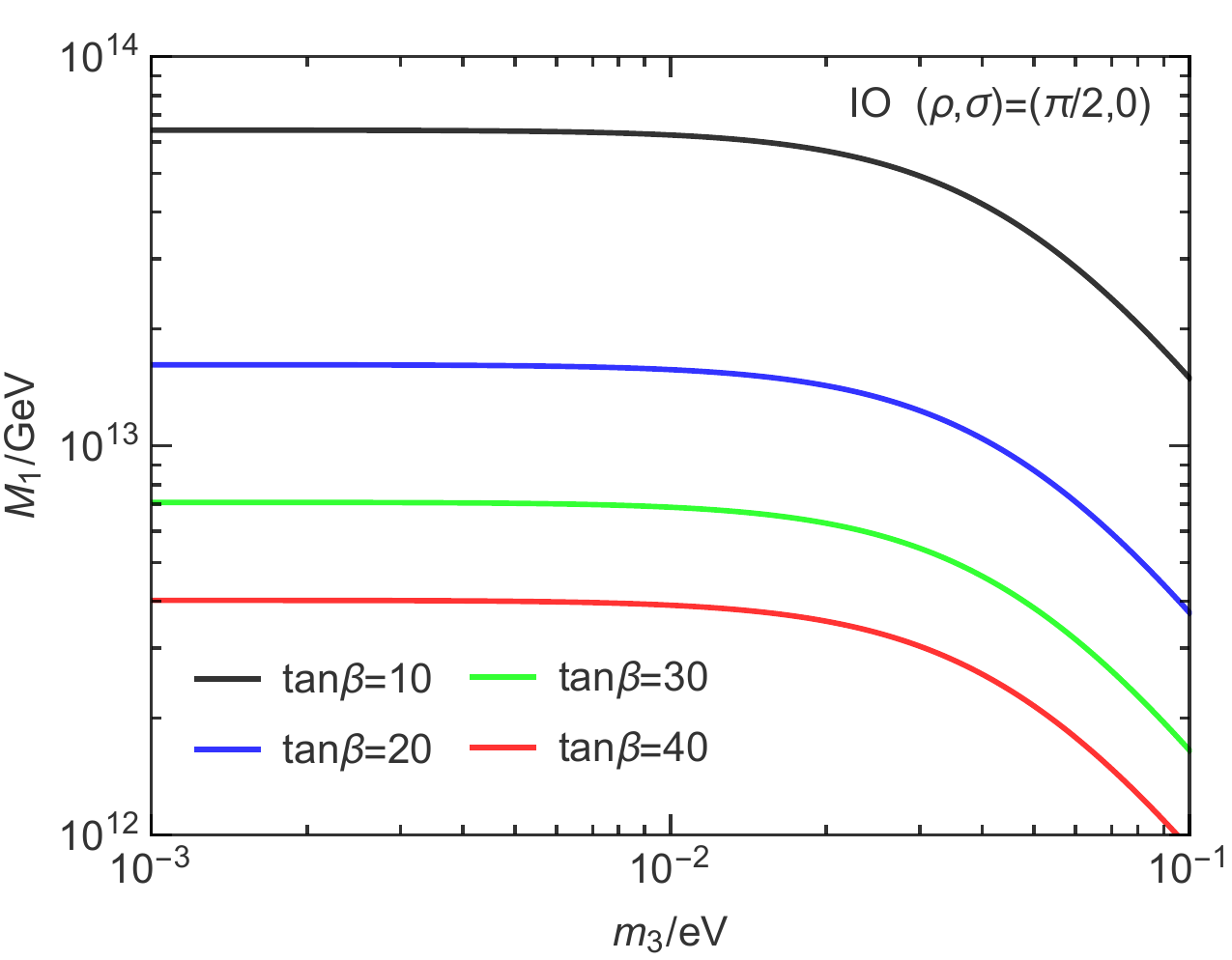}
\caption{ In the IO case, for some benchmark values of $\tan \beta$, the values of the lightest right-handed neutrino mass that allow for a reproduction of the observed value of $Y^{}_{\rm B}$ as functions of the lightest neutrino mass $m^{}_3$. }
\label{fig8}
\end{figure*}
%%%%%%%%%%%%%%%%%%%%%%%%%%%%%%%%%%%%%%%%%%%%%%%%%%

\section{Renormalization group evolution induced leptogenesis for low-scale leptogenesis}

It is well known that in the scenario of the right-handed neutrino masses being hierarchial there exists a lower bound about $10^9$ GeV for the right-handed neutrino mass scale to successfully reproduce the observed value of $Y^{}_{\rm B}$ \cite{DI}. Hence in such a scenario the right-handed neutrino mass scale is too high to be directly accessed by forseeable collider experiments. Fortunately, in the resonant leptogenesis scenario which is realized for nearly degenerate right-handed neutrinos, the CP asymmetries will get resonantly enhanced as \cite{resonant}
\begin{eqnarray}
\varepsilon^{}_{I\alpha} = \frac{{\rm Im}\left\{ (M^*_{\rm D})^{}_{\alpha I} (M^{}_{\rm D})^{}_{\alpha J}
\left[ M^{}_J (M^\dagger_{\rm D} M^{}_{\rm D})^{}_{IJ} + M^{}_I (M^\dagger_{\rm D} M^{}_{\rm D})^{}_{JI} \right] \right\} }{8\pi  v^2 (M^\dagger_{\rm D} M^{}_{\rm D})^{}_{II}} \cdot \frac{M^{}_I \Delta M^2_{IJ}}{(\Delta M^2_{IJ})^2 + M^2_I \Gamma^2_J} \;,
\label{5.1}
\end{eqnarray}
where $\Delta M^2_{IJ} \equiv M^2_I - M^2_J$ has been defined and $\Gamma^{}_J= (M^\dagger_{\rm D} M^{}_{\rm D})^{}_{JJ} M^{}_J/(8\pi v^2)$ is the decay rate of $N^{}_J$ (for $J \neq I$). Thanks to such a resonance enhancement effect, a successful leptogenesis can be realized even with TeV-scale right-handed neutrinos which have the potential to be directly accessed by running or upcoming experiments \cite{RHN}.
In this scenario, leptogenesis works in the three-flavor regime and the contributions of the two nearly degenerate right-handed neutrinos to the final baryon asymmetry will be on the same footing. Accordingly, the final baryon asymmetry is given by (taking $N^{}_1$ and $N^{}_2$ as the two nearly degenerate right-handed neutrinos as example, and similarly for the other cases of two right-handed neutrinos being degenerate) \cite{Lreview}
\begin{eqnarray}
Y^{}_{\rm B} =  c d \left[ (\varepsilon^{}_{1e} + \varepsilon^{}_{2e} ) \kappa ( \widetilde m^{}_{1e} + \widetilde m^{}_{2e}) +
(\varepsilon^{}_{1\mu} + \varepsilon^{}_{2\mu} ) \kappa ( \widetilde m^{}_{1\mu} + \widetilde m^{}_{2\mu}) + (\varepsilon^{}_{1\tau} + \varepsilon^{}_{2\tau} ) \kappa ( \widetilde m^{}_{1\tau} + \widetilde m^{}_{2\tau}) \right] \;.
\label{5.2}
\end{eqnarray}
Note that the efficiency factor for each flavor $\alpha$ is determined by the sum of the corresponding flavored washout mass parameters of the two nearly degenerate right-handed neutrinos.

For the model considered in the present paper, the resonant leptogenesis scenario may be realized given that in some parameter ranges there may exist two nearly degenerate right-handed neutrinos (see Figures~3 and 4). To be explicit, Figure~3 shows that in the NO case, for the benchmark value of $r=3$, there will be two nearly degenerate right-handed neutrinos at $m^{}_1 \simeq 0.005$ eV in the case of $(\rho, \sigma)=(0, 0)$, at $m^{}_1 \simeq 0.005$ eV and 0.01 eV in the case of $(\rho, \sigma)=(0, \pi/2)$, and at $m^{}_1 \simeq 0.006$ eV and 0.03 eV in the case of $(\rho, \sigma)=(\pi/2, 0)$. For the benchmark value of $r=10$, there will be two nearly degenerate right-handed neutrinos at $m^{}_1 \simeq 0.004$ eV and 0.007 eV in the case of $(\rho, \sigma)=(0, \pi/2)$, and at $m^{}_1 \simeq 0.002$ eV, 0.007 eV and 0.02 eV in the case of $(\rho, \sigma)=(\pi/2, 0)$. On the other hand, Figure~4 shows that in the IO case, for the benchmark value of $r=0.1$, there will be two nearly degenerate right-handed neutrinos at $m^{}_3 \simeq 0.002$ eV in the case of $(\rho, \sigma)=(0, 0)$, at $m^{}_3 \simeq 0.004$ eV in the case of $(\rho, \sigma)=(0, \pi/2)$, and at $m^{}_3 \simeq 0.02$ eV in the case of $(\rho, \sigma)=(\pi/2, 0)$. For the benchmark value of $r=0.3$, there will be two nearly degenerate right-handed neutrinos at $m^{}_3 \simeq 0.005$ eV in the case of $(\rho, \sigma)=(0, 0)$, at $m^{}_3 \simeq 0.01$ eV in the case of $(\rho, \sigma)=(0, \pi/2)$, and at $m^{}_3 \simeq 0.03$ eV in the case of $(\rho, \sigma)=(\pi/2, 0)$.

However, due to the relations in Eq.~(\ref{3.6}), leptogenesis cannot work when the $\mu$-$\tau$ reflection symmetry is exact. Fortunately, as mentioned in section~4, thanks to the difference between $y^{}_\mu$ and $y^{}_\tau$, the renormalization group evolution effects can break the $\mu$-$\tau$ reflection symmetry and consequently induce leptogenesis to work. Now we study the possibility of leptogenesis being induced by the renormalization group evolution effects for the resonant leptogenesis scenario. Taking account of the renormalization group evolution effects, the relations $\varepsilon^{}_{I\mu} =- \varepsilon^{}_{I\tau} $ and $\widetilde m^{}_{I\mu} = \widetilde m^{}_{I \tau}$ will be broken. To be explicit, one has
\begin{eqnarray}
&& \varepsilon^{}_{1 \mu} + \varepsilon^{}_{1\tau} = -\Delta^{}_\tau  \frac{\eta (\eta^{}_1 \eta^{}_2+1) M^2_0 r^2 (1-r^2) c^{}_x s^{}_x c^2_y s^{}_y }{16\pi v^2 [ c^2_x c^2_y + r^2 (s^2_x + c^2_x s^2_y) ] } \cdot \frac{M^2_1 \Delta M^2_{12}}{(\Delta M^2_{12})^2 + M^2_1 \Gamma^2_2} \; , \nonumber \\
&& \widetilde m^{}_{1 \tau} - \widetilde m^{}_{1\mu} = \Delta^{}_\tau \frac{M^2_0 r^2 (s^2_x + c^2_x s^2_y)}{M^{}_1} \;.
\label{5.3}
\end{eqnarray}
For the above-listed cases possessing two nearly degenerate right-handed neutrinos, we examine if the leptogenesis induced by the renormalization group evolution effects can successfully reproduce the observed value of $Y^{}_{\rm B}$. In the SM framework, it is found that only in the NO case and for the benchmark value of $r=3$ and $(\rho, \sigma)=(0, \pi/2)$ or $(\pi/2, 0)$ can the observed value of $Y^{}_{\rm B}$ be successfully reproduced. Figure~8 has shown the values of $\Delta M/M$ versus $M$ that allow for a reproduction of the observed value of $Y^{}_{\rm B}$. Here $M$ and $\Delta M$ are respectively the mass scale and splitting for the two nearly degenerate right-handed neutrinos.
These results are obtained by employing the ULYSSES package \cite{ulysses}. We see that, in order to give a successful leptogenesis, $M$ and $\Delta M/M$ (the degeneracy degree of the two nearly degenerate right-handed neutrinos) need to be in the ranges $10^4-10^5$ GeV and $10^{-14} - 10^{-12}$, respectively.
%%%%%%%%%%%%%%%%%%%%%% FIG 3%%%%%%%%%%%%%%%%%%%%%%
\begin{figure*}
\centering
\includegraphics[width=6.5in]{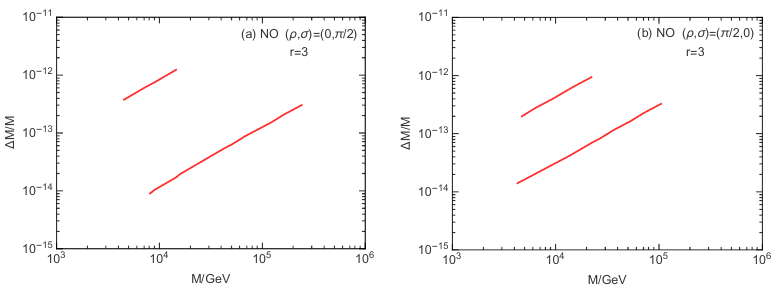}
\caption{ In the NO case, for the benchmark value of $r=3$, the values of $\Delta M/M$ versus $M$ that allow for a reproduction of the observed value of $Y^{}_{\rm B}$ in the cases of $(\rho, \sigma) =(0, \pi/2)$ and $(\pi/2, 0)$. }
\label{fig9}
\end{figure*}
%%%%%%%%%%%%%%%%%%%%%%%%%%%%%%%%%%%%%%%%%%%%%%%%%%

One may wonder how such a tiny right-handed neutrino mass splitting can be naturally realized and if it is stable against the renormalization correction. To kill these two birds with one stone, we examine the possibility that the required right-handed neutrino mass splitting is just generated by the renormalization correction itself. It is found that in the NO case and for $(\rho, \sigma)=(\pi/2, 0)$ this possibility can be viable. For this possibility, Figure~10 has shown the values of $|\Delta M|/M$ as functions of $M$ that allow for a reproduction of the observed value of $Y^{}_{\rm B}$.

%%%%%%%%%%%%%%%%%%%%%% FIG 3%%%%%%%%%%%%%%%%%%%%%%
\begin{figure*}
\centering
\includegraphics[width=5in]{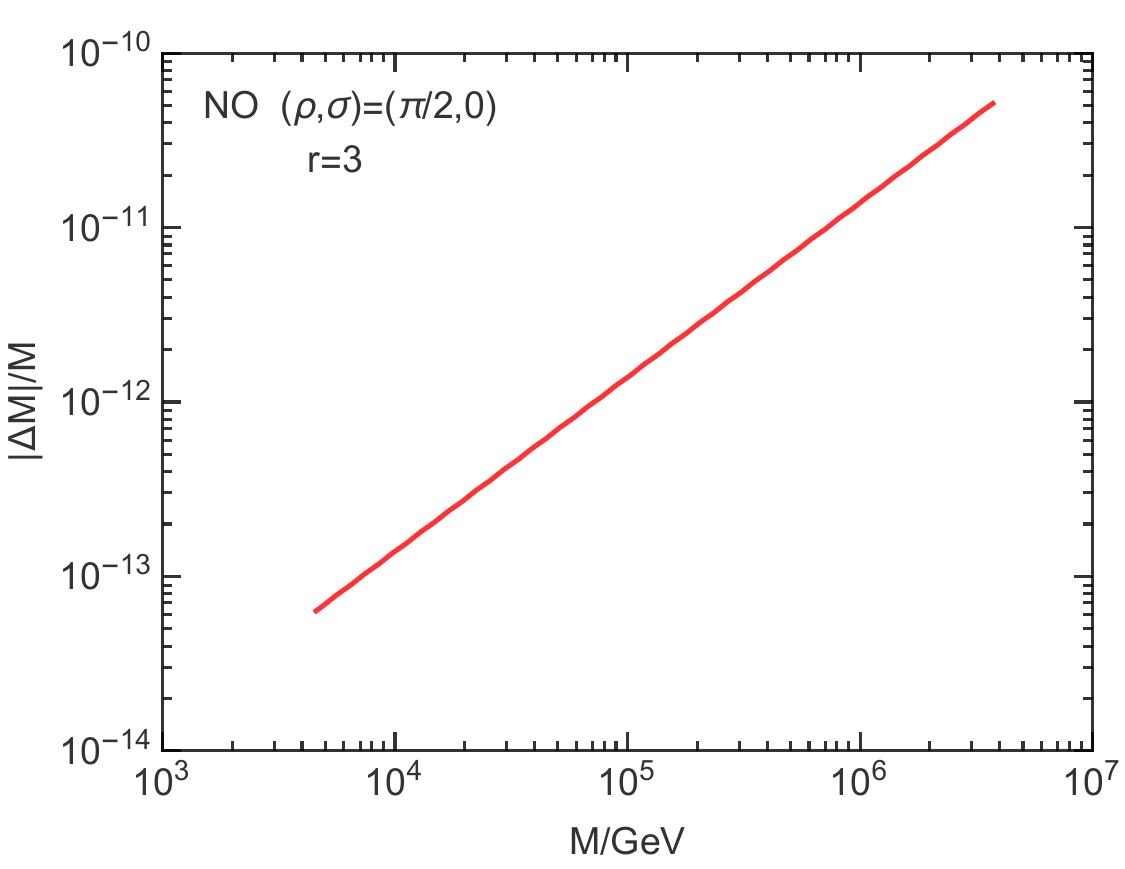}
\caption{ In the NO case with $(\delta, \rho, \sigma) =(3\pi/2, \pi/2, 0)$, for the benchmark value of $r=3$, the values of $|\Delta M|/M$ as functions of $M$ that allow for a reproduction of the observed value of $Y^{}_{\rm B}$. }
\label{fig9}
\end{figure*}
%%%%%%%%%%%%%%%%%%%%%%%%%%%%%%%%%%%%%%%%%%%%%%%%%%

\section{Summary}

Inspired by the special values of the neutrino mixing angles (in particular the closeness of $\theta^{}_{23}$ to $\pi/4$) and a preliminary experimental hint for $\delta \sim - \pi/2$, the $\mu$-$\tau$ reflection symmetry which can naturally predict $\theta^{}_{23} = \pi/4$ and $\delta= \pm \pi/2$ has been attracting a lot of attention. Furthermore, this symmetry also predicts the Majorana CP phases $\rho$ and $\sigma$ to be trivially 0 or $\pi/2$.
In this paper, we have studied the consequences of this symmetry for leptogenesis in the type-I seesaw model with the Dirac neutrino mass matrix being diagonal [see Eq.~(\ref{8})].

We have first derived the expressions of the model parameters $A$, $B$, $C$ and $D$ in Eq.~(\ref{8}) in terms of the neutrino mass and mixing parameters together with $M^{}_0$ and $r$ [see Eq.~(\ref{2.4})] and then shown their phenomenologically allowed values by taking account of the experimental results for the neutrino mass and mixing parameters and taking $r=1$ and $M^{}_0=1$ GeV as typical inputs (see Figures~1 and 2). The results show that there may exist zero or equal entries in $M^0_{\rm R}$ (which  in the literature are usually taken as smoking guns for some underlying flavor physics). Then, we have derived the dependence of three right-handed neutrino masses on $A$, $B$, $C$ and $D$ [see Eq.~(\ref{2.11})] and then shown their possible values as functions of the lightest neutrino mass $m^{}_1$ or $m^{}_3$ for some benchmark values of $r$ (see Figures~3 and 4). The results show that there may exist two nearly degenerate right-handed neutrinos in some parameter ranges, in which cases low-scale resonant leptogenesis can be realized.

Then, we have studied the consequences of the model considered in this paper for leptogenesis. Due to the $\mu$-$\tau$ reflection symmetry, leptogenesis can only work in the two-flavor regime. Furthermore, leptogenesis cannot work for the particular case of $r=1$, due to the orthogonality relations among different columns of $M^{}_{\rm D}$. Accordingly, for some benchmark values of $r \neq 1$, we have shown the values of the lightest right-handed neutrino mass $M^{}_1$ that allow for a reproduction of the observed value of $Y^{}_{\rm B}$ as functions of the lightest neutrino mass $m^{}_1$ or $m^{}_3$ in the cases of $(\rho, \sigma) = (0, 0), (0, \pi/2), (\pi/2, 0)$ and $(\pi/2, \pi/2)$ (see Figure~5). In the NO case, it is found that only in the cases of $(\rho, \sigma) = (0, 0)$ and $(\pi/2, \pi/2)$ and for $r>1$ can the observed value of $Y^{}_{\rm B}$ be successfully reproduced. And there exists an upper bound about 0.03 eV for $m^{}_1$. In the IO case, it turns out that only in the cases of $(\rho, \sigma) = (0, \pi/2)$ and $(\pi/2, 0)$ and for $r<1$ can the observed value of $Y^{}_{\rm B}$ be successfully reproduced. And there exists an upper bound about 0.1 eV or 0.02 eV for $m^{}_3$ in the case of $(\rho, \sigma) = (0, \pi/2)$ or $(\pi/2, 0)$.

Furthermore, we have investigated the possibilities of leptogenesis being induced by the renormalization group evolution effects for two particular scenarios. For the particular case of $r=1$, the renormalization group evolution effects will break the orthogonality relations among different columns of $M^{}_{\rm D}$ (due to the differences among $y^{}_\alpha$) and consequently induce leptogenesis to work. In the one-flavor leptogenesis regime, the  relevant CP asymmetry is suppressed by $\Delta^2_\tau$ [see Eq.~(\ref{4.6})], so the observed value of $Y^{}_{\rm B}$ cannot be successfully reproduced. In the two-flavor leptogenesis regime, the relevant flavored CP asymmetries are only suppressed by $\Delta^{}_\tau$ [see Eq.~(\ref{4.7})]. In the SM framework, due to the smallness of $\Delta^{}_\tau$, the resultant $Y^{}_{\rm B}$ cannot reproduce its observed value. But in the MSSM framework the observed value of $Y^{}_{\rm B}$ may be successfully reproduced due to the enhancement effects from a large $\tan \beta$ value. For some benchmark values of $\tan \beta$, we have shown the values of the lightest right-handed neutrino mass $M^{}_1$ that allow for a reproduction of the observed value of $Y^{}_{\rm B}$ as functions of the lightest neutrino mass $m^{}_1$ or $m^{}_3$ in the phenomenologically allowed cases of $(\rho, \sigma)$ (see Figure~7).

For the low-scale resonant leptogenesis scenario which is realized for nearly degenerate right-handed neutrinos, leptogenesis cannot work when the $\mu$-$\tau$ reflection symmetry is exact. Similarly, thanks to the difference between $y^{}_\mu$ and $y^{}_\tau$, the renormalization group evolution effects can break the $\mu$-$\tau$ reflection symmetry and consequently induce leptogenesis to work.

\vspace{0.5cm}

\underline{Acknowledgments} \vspace{0.2cm}

This work was supported in part by the National Natural Science Foundation of China under grant Nos.~11605081, 12142507 and 12147214, and the Natural Science Foundation of the Liaoning Scientific Committee under grant NO.~2022-MS-314.

\end{document}